\documentclass[aps,prep,twocolumn,preprintnumbers,floatfix,nofootinbib,showpacs]{revtex4}

\usepackage{epsfig,subfigure,graphicx,amsmath,amssymb}
\usepackage{color}
\usepackage{mathrsfs}
\usepackage{verbatim}
\usepackage{mdwlist}

\newcommand{\beq}{\begin{equation}}
\newcommand{\eeq}{\end{equation}}
\newcommand{\bea}{\begin{eqnarray}}
\newcommand{\eea}{\end{eqnarray}}

\newcommand{\TeV}{\,\mathrm{TeV}}
\newcommand{\GeV}{\,\mathrm{GeV}}

\newcommand{\pb}{\,\mathrm{pb}}

\usepackage{ulem}
\def\etmiss{{E\hskip -0.2cm\slash \hskip 0.01cm }_T}

\def\squark{\tilde{q}}
\def\gluino{\tilde{g}}
\def\chargino{\tilde{\chi}_1^{\pm}}
\def\neutralino2{\tilde{\chi}_2^0}
\def\slepton{\tilde{l}}
\def\stau{\tilde{\tau}}

\def\sneutrino{\tilde{\nu}}
\def\lnp{\tilde{\chi}_1^0}

\def\dt{\delta_{T}}

\def\absdtbar{|\bar{\delta}_{T}|}

\def\alt{\alpha_{T}}

\def\bet{\beta_{T}}

\def\kat{k_{1T}}

\def\kbt{k_{2T}}

\def\mmt{M_{T}}
\def\mt2{M_{T2}}
\def\rmt2{M_{T2R}}
\def\lmt2{M_{T2L}}
\def\cmt2{Combinatoric M_{T2}}
\def\trialmt2{M_{T2}(\chi)}
\def\trialcmt2{Combinatoric\,M_{T2}(\chi)}
\def\BK2{^{2}BK}

\def\mmct{M_{CT}}
\def\mct2{M_{CT2}}
\def\lmct2{M_{CT2L}}
\def\rmct2{M_{CT2R}}
\def\cmct2{Combinatoric\,M_{CT2}}
\def\trialmct2{M_{CT2}(\chi)}
\def\trialcmct2{Combinatoric\,M_{CT2}(\chi)}

\def\mctt2{M_{CT2,T2}}

\def\p0{{\bf p}_0}

\begin{document}

\preprint{IPMU 10-130}

\title{Mass measurement in boosted decay systems at hadron colliders}

\author{Won Sang Cho$^{a}$}
\author{William Klemm$^{a,b}$}
\author{Mihoko M. Nojiri$^{a,c,d}$}
\vspace{0.4cm} \affiliation{${}^a$ Institute for the Physics and
Mathematics of the Universe, University of Tokyo, Chiba 277-8582,
Japan} \affiliation{${}^b$ Department of Physics, University of
California, Berkeley, California 94720, USA} \affiliation{${}^c$
Theory Group, KEK, 1-1 Oho, Tsukuba, Ibaraki 305-0801, Japan}
\affiliation{${}^d$ The Graduate University for Advanced Studies
(SOKENDAI),1-1 Oho, Tsukuba, Ibaraki 305-0801, Japan}

\begin{abstract}
We report a new possibility of using the $\mct2$ (Constransverse
mass) variable for mass measurement in single step decay chains
involving missing particles with moderate transverse momentum. We
show that its experimental feasibility is enhanced compared to the
corresponding $\mt2$-kink method. We apply this method to
reconstruct a pair of chargino decay chains.
\end{abstract}

\pacs{12.60.Jv, 13.85.Hd, 14.80.Ly, 13.90.+i }

\keywords{LHC. Missing energy, $\mct2, $\mt2$, Transverse mass kink}

\maketitle

%%%%%%%%%%%%%%%%%%%%%%%%%%%%%%%%%%%%%%%%%%%%%%%%%%%%%%%%%%%%%%%%%%%%%%%%%%%%

Many theories of new physics beyond the Standard Model (SM) are
expected to provide a rich invisible energy signal from their
Lightest New Particles (LNP), which are stable Dark Matter
candidates, missing in the detector of future hadron colliders. In
this situation only the so-called `$\mt2$-kink' method can provide
the information necessary to determine the masses of both mother
particle and missing LNP simultaneously for the events with a pair
of single step decay chains \cite{mt2masskink,mt2boostkink}.

The origin of the $\mt2$-kink is the variety of `Extreme Kinematic
Configurations (EKCs)' in the events which can contribute to the
maximum of the $\mt2$ distribution. In general, for a given trial
LNP mass, $\chi$, the different EKC's will take on different values,
so in different regions of $\chi$, $\mt2^{\textrm{max}}(\chi)$ will
follow different functional forms. At the true value, $\chi = m_X$,
all of the EKC should provide the same $\mt2$ maximum value as the
true mother particle mass, $m_Y$, by the definition of $\mt2$
\cite{mt2}. Consequently, $\mt2^{\textrm{max}}(\chi)$ shows a slope
discontinuity at the kink point,
$\mt2^{\textrm{max}}(\chi=m_X)=m_Y$.

Let us consider the system of a pair of single step decay chains at
the LHC:
%\begin{eqnarray}
$p + p \rightarrow \dt + Y_{1}/Y_{2}( \rightarrow \alpha X_{1}/\beta
X_{2}),\label{lhcevents}$
%\end{eqnarray}
where $Y_{1,2}$ are the two mother particles with identical masses,
each decaying to visible $\alpha(\beta)$ and missing LNP
($X_{1(2)}$). $\dt$ denotes the other remnants (and the sum of their
transverse momenta) not from $Y_{i}$ decays, i.e. from the Initial
State Radiation (ISR) or decays before $Y_{i}$, so that they provide
the total transverse momentum of the $Y_{1+2}$ system of $-\dt$. In
this event topology, there are two physical degrees of freedom that
can generate the variety of EKC developing the kink. One is the
invariant mass of visible particles $m_{\alpha,\beta}$ in
$N$($\geq3$)-body decays, generating a so-called `Mass Kink'
\cite{mt2masskink}, and the other one is non-zero transverse
momentum ($\dt$) leading to a `Boosted Kink (BK)'
\cite{mt2boostkink} of the $Y_{1+2}$ system of our interest. When
$\alpha(\beta)$ consists of a single visible particle and has a
fixed invariant mass, then only the BK provides enough constraints
for simultaneous mass measurement.

However, the BK is not easy to identify \cite{mt2boostkink}. It
requires large $\dt$ to have a clear kink structure. With the
highest practical values of $\dt/m_Y\sim O(1-10)$ at the LHC, the
kink structure is not clear enough to be reliably measured. The
profile of the $\dt$ distribution is sharply decreasing for large
values at the LHC, and the endpoint structure of the $\mt2(\chi)$
distribution is smeared for large $\dt$ with small statistics. This
can introduce significant systematic errors in the endpoint
extractions.

In this paper we present the endpoint behavior of the $\mct2(\chi)$
distribution \cite{mct2} for single step decay chains with non-zero
$\dt$. The $\mct2^{\textrm{max}}(\chi)$ distributions for different
$\dt$ carry independent information about $m_Y$ and $m_X$, and it
can be utilized for simultaneous mass measurement of mother and
missing particles. Furthermore, as the dependence of $\mct2$ on
$\dt$ is much more significant compared with the $\mt2$, it becomes
good collider basis in resolving the boundary of the distorted phase
space by various transverse boost effects. One more interesting
property is that the endpoint structure is statistically cleaner
with much less pollution from unwanted width effect in addition to
the amplification and sharpening of the structure as noticed in
\cite{mct2}. In next sections, we describe the merit of using
$\mct2^{\textrm{max}}$ over $\mt2^{\textrm{max}}$ for systems in
which reasonable $\dt$ is expected from heavy colored NP particle
decays. We demonstrate that it can be used to probe the boosted
kinetic boundary, and determine the masses of the particles involved
in the chargino decays into sleptons, in particular $\chargino
\rightarrow \sneutrino(\rightarrow\lnp+\nu) +\ell$.

The $\mct2$ (Constransverse mass) variable \cite{mct2} for the
$Y_{1+2}$ system is defined by replacing $\mmt$ with $\mmct$
\cite{mct} in the definition of $\mt2$ \cite{mt2} as follows:
\begin{eqnarray}
\mct2(\chi) &=& \min_{\kat + \kbt =\etmiss}
\left[\max{\{\mmct^{(1)},\mmct^{(2)}\}}\right]\label{mct2def}\\
\mmct^{(1)}(\chi)^{2} &=& \chi^2 + 2 (|\alt| e_{1} + \alt \cdot
\kat),\nonumber
\end{eqnarray}
where $\mmct^{(2)}(\chi)$ is defined by replacing
($\alt,e_1,k_{1T}$) with ($\bet,e_2,k_{2T}$). $k_{1,2T}$ and
$e_{1,2}$ are the transverse (Tr) momenta and Tr energy,
respectively, of missing $X_{1,2}$, with total missing Tr momentum,
$\etmiss=-(\alt+\bet)-\dt$. $\alt(\bet)$ represents the Tr momenta
of visible particles from $Y_{1}(Y_{2})$ decays, $-\dt$ is the total
Tr momentum of the $Y_{1+2}$ system, and $\chi$ denotes the trial
test mass of $X_{i}$. Visible particle masses are assumed to be
zero, however the definition of $\mct2$ can be easily generalized
for massive visible particle events. In \cite{mct2}, it was found
that the number of events near $\mct2^{\textrm{max}}$ can be
amplified with a judicious choice of trial mass, $\chi$, and $\mct2$
was employed to measure some physical constraints involving squark
and gluino masses with fewer systematic errors in pinpointing the
endpoint than with the $\mt2$ analysis. One thing worthy to note
here is that interestingly, the endpoint of the $\mt2(\chi)$
\cite{mt2masskink,mt2boostkink} or $\mct2(\chi)$ \cite{mct2} for any
$\chi$ values, is invariant under the unknown $\sqrt{s}(\geq 2 m_Y)$
values of the system with two mother particles, and this property is
also satisfied by the $\mct2^{\textrm{max}}(\chi)$ for general
boosted decay system as in following paragraphs.

When $\dt=0$, the $\mct2(\chi)$ has a one-to-one correspondence with
$\mt2(\chi)$, and its maximum value is given as follows \cite{mct2}
:
\begin{eqnarray}
\mct2^{max 2}(\chi) = \chi^2 + 2 (\alpha_0 E_{X}^0
-\alpha_0^2),\label{mct2max}
\end{eqnarray}
where $\alpha_0=\frac{m_Y^2-m_X^2}{2 m_Y}$ and
$E_{X}^0=\sqrt{\chi^2+\alpha_0^2}$. However, if $\dt\neq0$, simple
one-to-one correspondence no longer exists, except for the events of
some EKCs. Certainly, depending on $\chi$, $\mct2(\chi)$ and
$\mt2(\chi)$ can be one-to-one, however, it's not valid in general
$\chi$ values. In result, the maximum profile of the $\mct2(\chi)$
distribution shows a 2nd order Boosted Kink, ($^2$BK), structure
which is different from the BK of $\mt2$. As investigated in
\cite{mt2masskink,mt2boostkink}, the $\mt2^{max}(\chi,\dt)$ value of
the $Y_{1+2}$ system is the same as the $\mmt^{max}(\chi,\dt/2)$ of
single $Y_i$ decay system, and the EKC for the $\mt2^{max}$ can be
characterized by a pair of identical EKCs corresponding to the
$\mmt^{max}$ of single $Y_i$ decays although they have to be
combined to produce the event with general $\sqrt{s}\geq2 m_Y$.
Then, in the language of single $Y_i$ decay, the BK of
$\mt2^{max}(\chi)$ is provided by two EKCs of the single $Y_i$ decay
events, characterized as follows :
\begin{eqnarray}
\textrm{\textbf a)~} \phi^{max}=0 \textrm{~for~} \chi \leq m_X
\textrm{~~~\textbf b)~} \phi^{max}=\pi \textrm{~for~} \chi \geq
m_X\nonumber
\end{eqnarray}
where $\phi^{max}$ is the azimuthal angle between visible Tr
momentum and $\dt$ in the rest frame of the $Y_{i}$ with vanishing
$\Delta\eta(\equiv\eta_{vis}-\eta_{inv})$. Similarly, the EKCs for
$\mct2^{max}(\chi)$ with $^2$BK can be also characterized by single
$Y_i$ decays, but the azimuthal angle dependence is different :
\begin{description}
    \item[1)] $\cos\phi^{max}=\frac{m_X^2 m_Y^2}{\absdtbar E_Y (m_Y^2-m_X^2)}
    \left[\frac{\chi^2}{m_X^2}+\frac{\absdtbar^2}{m_Y^2}\right]$
    for $\chi \leq \chi_*$
    \item[2)] $\cos\phi^{max}=-1$ for $\chi\geq \chi_*$,
\end{description}
where $\absdtbar=|\dt|/2$ and $E_Y=\sqrt{m_Y^2+\absdtbar^2}$. Thus,
$\mct2^{\textrm{max}}(\chi)$ is described by two curves in the two
regions of $\chi$ divided at $\chi_*$, where $\chi_*$ is given by
\begin{eqnarray}
\chi_*^2=\absdtbar\left(2\alpha-\absdtbar\right), \label{2ndkink}
\end{eqnarray}
with
$\alpha=|\alpha_0|(\frac{\absdtbar}{m_{Y}}+\frac{E_{Y}}{m_{Y}})$.
Then $\mct2^{\textrm{max}}$ for the full $\chi$ range is given as
follows :
\begin{eqnarray}
\mct2^{\textrm{max} 2} &=& 2\chi^2 + \absdtbar^{2} \,\,
\textrm{~~~~~~~~~~~~for $\chi\leq\chi_*$}\label{mct2left}\\
&=&\chi^2 + 2\alpha(\absdtbar-\alpha) + 2\alpha \sqrt{\chi^2 +
(\absdtbar-\alpha)^2} \nonumber\\
& &\qquad\textrm{~~~~~~~~~~~~~~~~~~~~~~for
$\chi\geq\chi_*$}.\label{mct2right}
\end{eqnarray}
When $\chi_*$ is real, the two maximal curves of Eq.
(\ref{mct2left}) and (\ref{mct2right}) come into contact at $\chi_*$
with the same inclination, but the curvature of
$\mct2^{\textrm{max}}(\chi)$ has a discontinuity at $\chi=\chi_*$
which we denote $^2$BK. It's interesting that when a $^2$BK exists,
there is a boost-trapped distribution for $\chi<\chi_*$ in that its
boundary is independent of any physical masses in the decay system,
as shown in Eq. (\ref{mct2left}). If $\chi_*$ is imaginary,
$\mct2^{\textrm{max}}(\chi\geq0)$ just follows the line given by Eq.
(\ref{mct2right}). For given parameters, ($m_Y,m_X,|\dt|$), $\chi_*$
is real and observable if
\begin{enumerate}
    \item $\frac{m_{X}}{m_{Y}}\leq\frac{1}{\sqrt{2}}$,
    then $\chi_* \in \bf{R}$.
    \item $\frac{1}{\sqrt{2}}<\frac{m_{X}}{m_{Y}}<1$,
    then $\chi_* \in \bf{R}$ if $|\dt|\leq|\dt^{*}|$, \\
    where $|\dt^{*}|\equiv 4 \alpha_0/\sqrt{1-\frac{4\alpha_0}{m_{Y}}}$.
\end{enumerate}
\begin{figure}[t!]
\begin{center}
\includegraphics[width=0.5\textwidth]{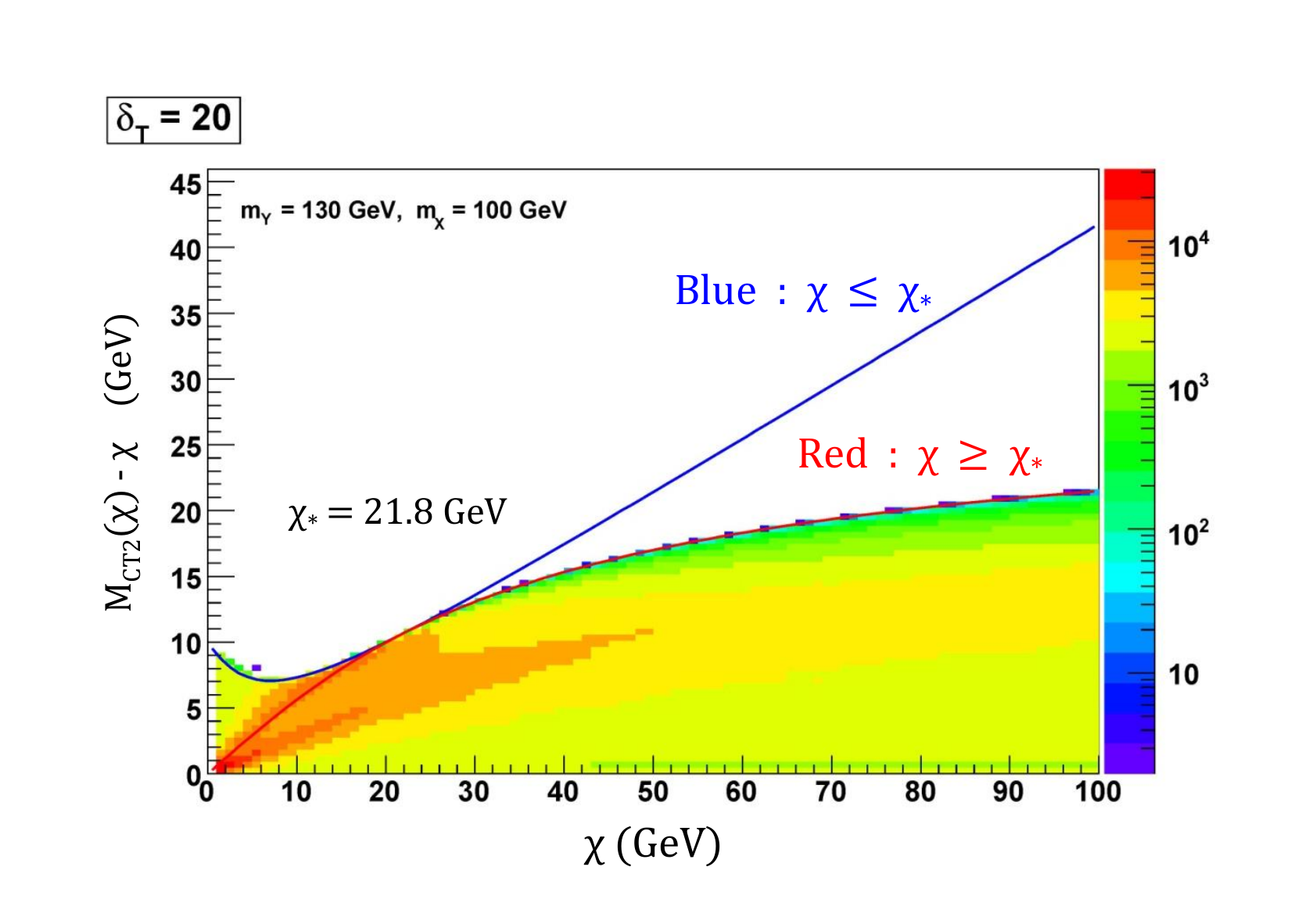}
\includegraphics[width=0.5\textwidth]{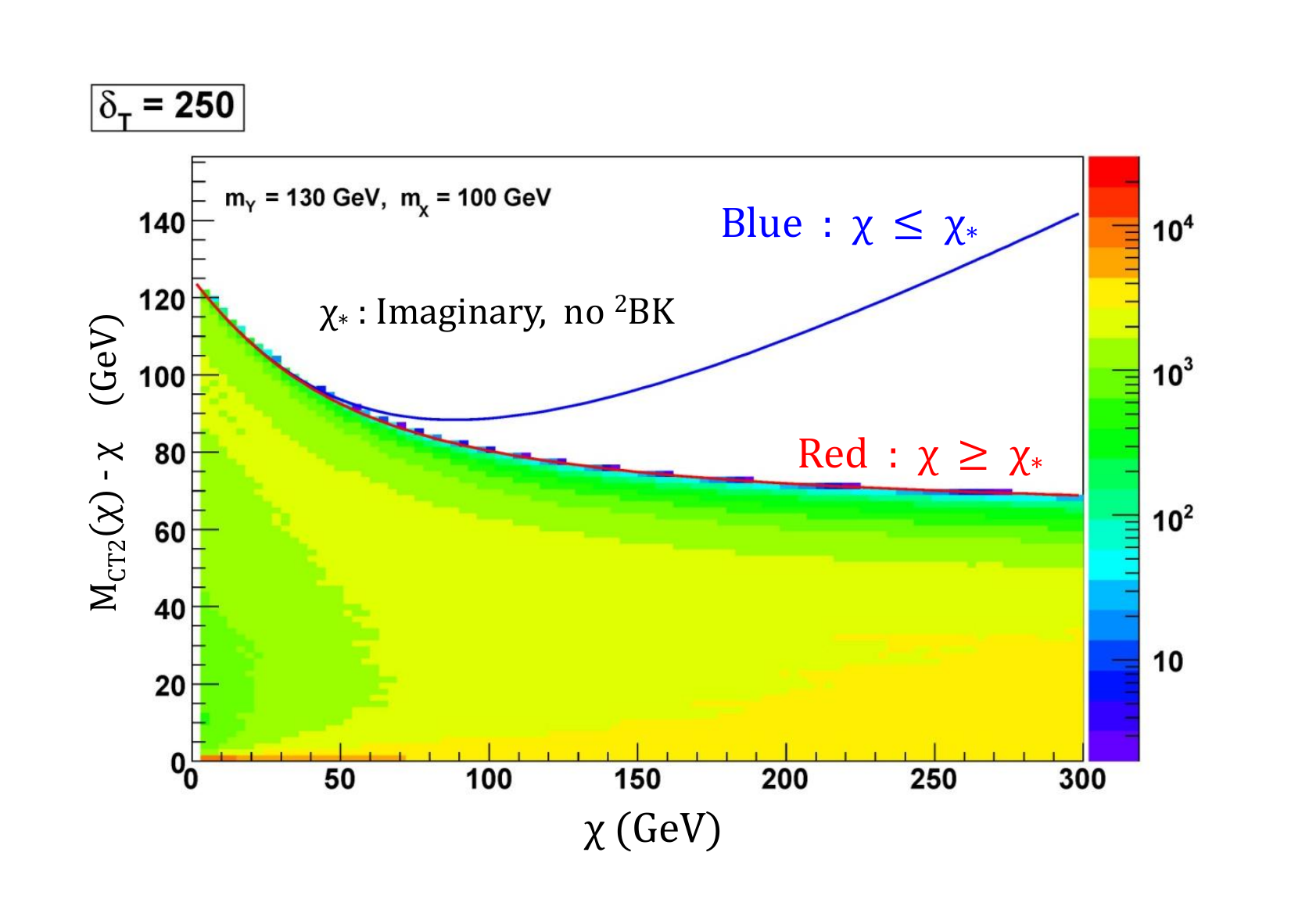}
\caption{a) $\mct2^{max}(\chi)$ for $(m_Y,m_X)=(130,100)\GeV$ and
$\delta_T^{*}=247.9\GeV$ with $\dt=20\GeV$ ($\chi_*=21.8\GeV$) b)
$\mct2(\chi)$ distribution for $\dt=250\GeV$ (No $^2$BK).
}\label{fig:mct2max}
\end{center}
\end{figure}
The reality condition of $\chi_*$ for the existence of a $^2$BK is
not always met, however, it certainly appears for the events with
small values of $|\dt|$. For example, Fig. (\ref{fig:mct2max}) shows
contour histogramming on the $\mct2(\chi)-\chi~\rm{vs}~\chi$ plane.
We used the $10^5$ Monte-Carlo events which simulate the phase space
of the pair of boosted decay system, $Y_{1+2}+\dt$ where
$(m_Y,m_X,\dt)=(130,100,20)\GeV$ with massless visible particle. In
Fig. (\ref{fig:mct2max}(a)) the $\mct2^{\textrm{max}}(\chi)$ is well
described by the two curves in the two regions of $\chi$ divided by
$^2$BK, $\chi_*=21.8\GeV$. For $\chi\leq\chi_*$,
$\mct2^{\textrm{max}}(\chi)$ follows the blue curve; it switches to
the red one for $\chi\geq\chi_*$ with a continuous slope at
$\chi=\chi_*$. Since the spectrum for Fig. (\ref{fig:mct2max})
belongs to the second category of reality condition with
$\dt^{*}=247.9 \GeV$, no $^2$BK arises for $\dt=250 \GeV > \dt^{*}$
as shown in Fig. (\ref{fig:mct2max}(b)). In any case one can try to
fit the endpoints with the segmented maximal endpoint function, Eq.
(\ref{mct2left}) and (\ref{mct2right}), for given maximal value of
$|\dt|$, $|\dt^{\textrm{max}}|$ in an event sample, since the
$\mct2^{\textrm{max}}(\chi)$ always increases with $|\dt|$. The
segmentation point, $\chi_*$ of the piecewise fitting function, Eq.
(\ref{mct2left}) and (\ref{mct2right}) is also a function of
$\alpha$, and overall fitting procedure systematically fits on
$\alpha$ without a prior knowledge about $\chi_*$. Then for several
$|\dt^{\textrm{max}}|$ values and given $\chi$ values, one can
extract independent mass constraints,
$\alpha(m_Y,m_X,|\dt^{\textrm{max}}|)$ by which the $m_Y$ and $m_X$
can be simultaneously measured.

The $\mct2$ is found to be so sensitive under the Tr. boost of the
system that even with moderate values of non-zero $\dt$, it can
provide good resolving power in discrimination of various transverse
boost effects which are usually buried in the large systematic
errors of shifted endpoint extraction. The rates of
$M_{CT2/T2}^{max}(\chi)$ shift with respect to a change in $\dt$ are
given as follows :
\begin{eqnarray}
&&\frac{\partial M_{CT2R/T2R}^{\textrm{max}}}{\partial
\absdtbar}\label{mct2t2shift}\\
&&~=\frac{\alpha E_{\chi}}{M_{CT2R/T2R}^{\textrm{max}}E_Y}
\left\{1\pm\frac{\absdtbar-\alpha}{E_{\chi}}\right\}
\left\{1\pm\frac{E_Y-\alpha}{E_{\chi}}\right\}\nonumber,
\end{eqnarray}
 for $M_{CT2/T2}$, respectively, with
$E_{\chi}\equiv\sqrt{\chi^2+(\absdtbar-\alpha)^2}$ and
$M_{T2R}^{\textrm{max2}}=\chi^2 - 2\alpha(\absdtbar-\alpha) +
2\alpha E_{\chi}$. Here the subscript letter-$R$ denotes the values
of $M_{CT2/T2}(\chi)$ for the $\chi$ values larger than
corresponding kink positions, $\chi_*$/$m_X$. The shift of the
$M_{CT2}^{\textrm{max}}$ is magnified in compared with
$M_{T2}^{\textrm{max}}$ as indicated from the flipped signs between
the rates of the endpoint shifts in Eq. (\ref{mct2t2shift}). In
particular, the shift can be large by the factor of
\,O(10)\footnote{Actually, the sensitivity is not only for the EKC
events. The relation (\ref{mct2t2shift}) holds for all events with
general visible transverse momentum smaller than $\alpha$.}, when
the mass difference between $M_Y$ and $M_X$ is small enough so that
$\absdtbar>\alpha$ for a moderate value of $\absdtbar$.
\begin{figure}[ht!]
\begin{center}
\includegraphics[width=0.4\textwidth]{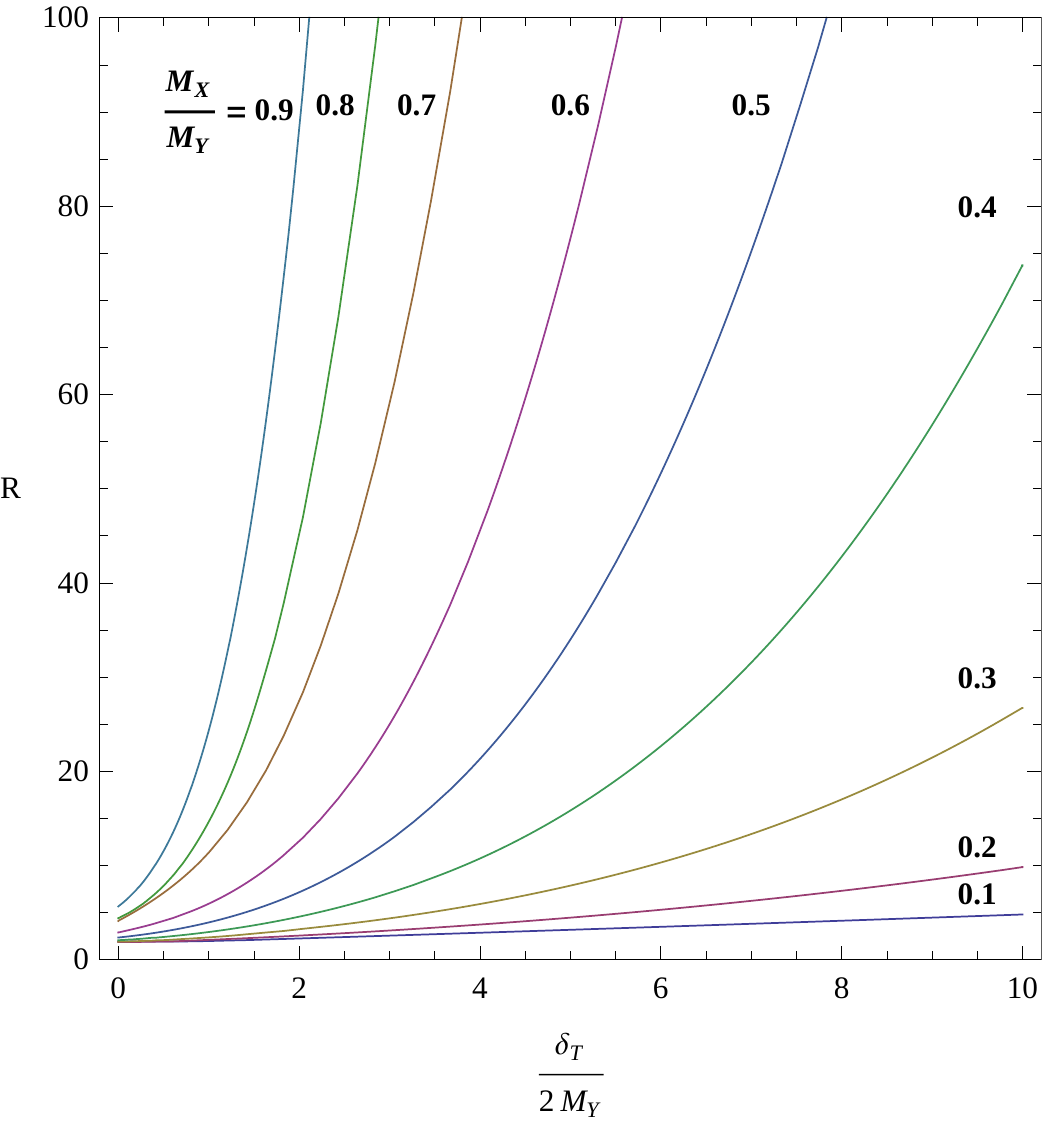}
\includegraphics[width=0.4\textwidth]{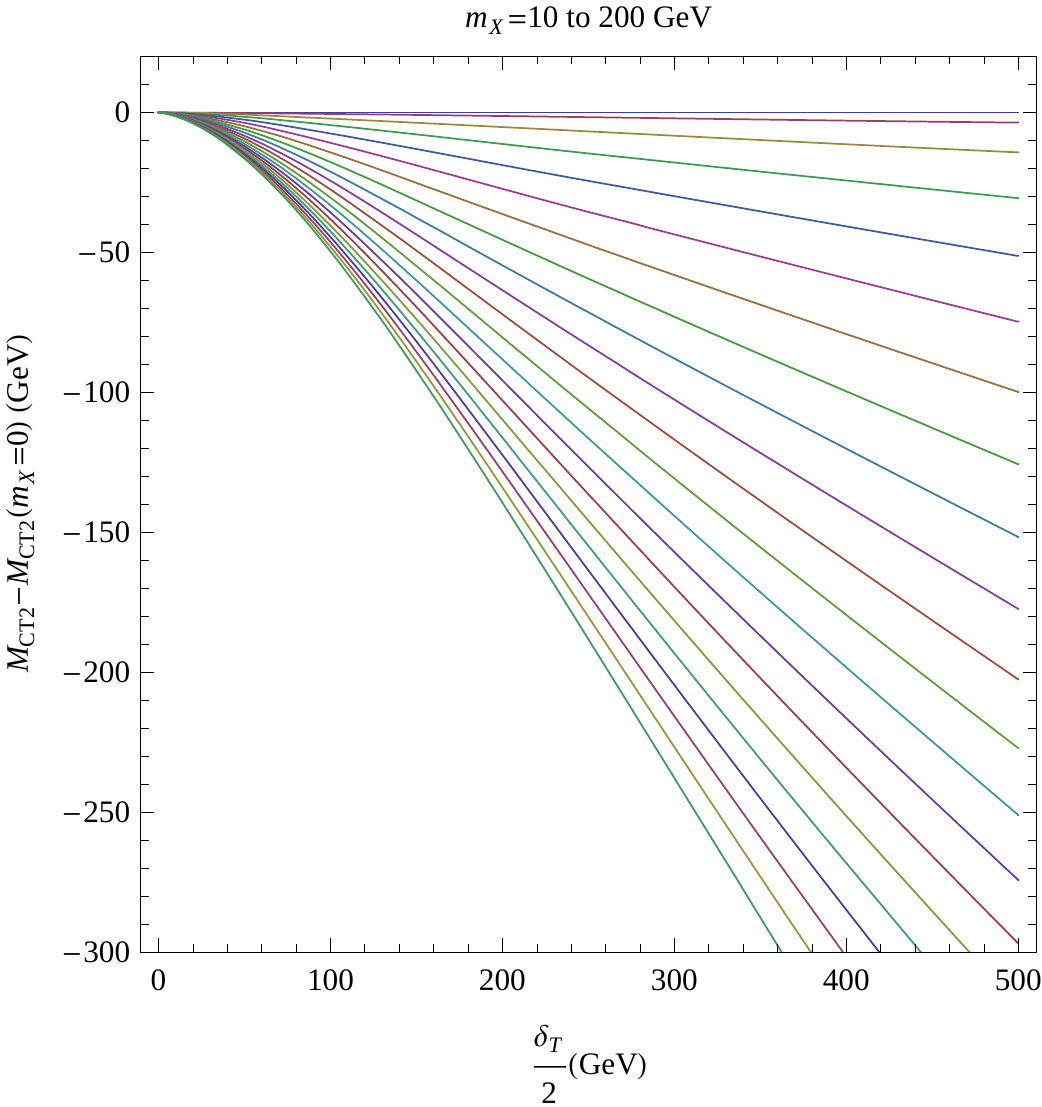}
\caption{ a) Ratios of $\dt$-shift between $\mct2^{\textrm{max}}$
and $\mt2^{\textrm{max}}$ with respect to $\dt/M_Y$ for various mass
spectrums, $M_X/M_Y$. b) $\dt$-shift of
$\mct2^{\textrm{max}}-\mct2^{\textrm{max}}(m_X=0)$ for
 $m_X=(10-190)\GeV$ with the $\alpha_0=62.4 \GeV$ and
the trial LNP mass, $\chi=\chi_*+50 \GeV$.}\label{fig:shift}
\end{center}
\end{figure} In Fig. (\ref{fig:shift}(a)) we show the shift of the
$\mct2^{\textrm{max}}$ compared with that of $\mt2^{\textrm{max}}$
under the effect of non-zero $\dt$. We define $R$ as
\begin{eqnarray}
R(\frac{\absdtbar}{m_Y},\frac{m_X}{m_Y})
\equiv\frac{\mct2^{\textrm{max}}(\chi_1,\absdtbar+\Delta)
-\mct2^{\textrm{max}}(\chi_1,\absdtbar)}
{\mt2^{\textrm{max}}(\chi_2,\absdtbar+\Delta)-\mt2^{\textrm{max}}(\chi_2,\absdtbar)}.\nonumber
\end{eqnarray}
We take $\Delta=m_Y/2$. Also, if $m_X/m_Y=0.1-0.7$, $\chi_1=2
\chi_*$, where $\chi_{*}$ is calculated with
$\absdtbar=\absdtbar+\Delta$, while $\chi_1=m_X+m_Y/2$ if
$m_X/m_Y=0.8-0.9$. Under the transverse boost effect, the magnified
shift of $\mct2^{\textrm{max}}(\chi,\dt)$ can be clearly seen
depending on the various mass ratios of $m_X/m_Y$ in Fig.
(\ref{fig:shift}(a)). If we compare the shifts of the
$M_{CT2/T2}^{\textrm{max}}$ using specific mass spectrums with
transverse boost effect,
$(m_Y,m_X,\absdtbar,\Delta)=(150,100,100,100)|(500,100,100,100)$,
then the shifts of $M_{CT2/T2}^{\textrm{max}}(\chi)$ are calculated
to be $\Delta\mct2^{\textrm{max}}(\chi=\chi_*)=73.5|38.4 \GeV$ while
$\Delta\mt2^{\textrm{max}}(\chi=200 \GeV)=12.7|11.6 \GeV$.

In calculating the $R$, the optimal choices of $\chi_{1,2}$ for
which each of the shifts is maximized and well-measured, are
ambiguous. We choose $\chi_1\sim O(\chi_*)$ because it always
provides sizable shift with clean and sharp endpoint structures of
the $\mct2$ distributions \cite{cmct2}. A practical choice of
$\chi_2$ is also unclear. Indeed, as we take larger value of
$\chi_2$, the denominator of $R$ increases, reducing $R$ down to
$O(1)$ in some cases. However, as we will see in next section, there
appear faint long tails near the expected endpoints of the $\mt2$
distributions for large $\dt$, making it difficult to measure the
endpoints and the shifts. Also, for a given value of $\dt$, such an
endpoint structure becomes worse as the $\chi$ increases over the
true value of $\chi=m_X$, while the denominator of $R$ gets rapidly
saturated just with $\chi\sim O(m_X)>m_X$. Thus for the sake of both
of the endpoint sharpness and the sizable endpoint shift, we choose
$\chi_2-m_X\sim m_Y/2$ so as to fix the separation of $\chi$ from
the true mass, $m_X$ to be the order of $m_X$, at least to be
$m_X/2$ in Fig. (\ref{fig:shift}(a)).

Fig. (\ref{fig:shift}(b)) displays the $\dt$-shift of
$\mct2^{\textrm{max}}-\mct2^{\textrm{max}}(m_X=0)$ for various
missing LNP masses, $m_X=(10-190)\GeV$ while the $\alpha_0$ is fixed
by $62.4 \GeV$ and the trial LNP mass, $\chi=\chi_*+50 \GeV$. For
$\dt=0$, no mass resolving power is obtained by the endpoint
measurement as expected, however, one can see that the $\dt$-shift
of $\mct2^{\textrm{max}}$ can be large enough to measure the new
particle masses with the resolution of $O(1-10)\GeV$ for $\dt\sim
O(10-100)\GeV$. It is also enhanced for the large $m_X/m_Y$ case. We
fixed the $\alpha_0$ because it is the basic momentum scale one can
observe in detector regardless of the mass spectra.

By projecting events in the $\mct2(\chi\sim O( \chi_*),\dt\neq0)$
basis, one can get a more sharper edged and $\dt$-sensitive event
distribution. The flipped sign in the definition of the $\mct2$
variable makes the distribution compact with respect to the internal
momentum scale of the system, while being much more sensitive for
the external boost momentum $\dt$, like a rubber ball. This means
that one can have a better chance to measure both of the masses with
enhanced resolution in probing the boosted kinematic boundaries.

In the following example, we employ the $\mct2(\chi)$ variable for
resolving superparticle masses, especially in the events with a pair
of $\chargino$ decays. It is well-known that one can efficiently
select such NP events by imposing the same-sign (SS) dileptonic
condition \cite{atlas2}. This signature provides an important
discovery channel in establishing SUSY, because the background is
small, however, mass measurement in this channel is difficult
because of the existence of extra missing $\nu$s and combinatorial
SUSY backgrounds of
$\chargino(\rightarrow\sneutrino+\ell,\slepton+\nu,W+\lnp)$ decays.
There exist a few studies \cite{ssboost} on mass measurement in this
event topology. We demonstrate that the mass resolving power is
significantly enhanced by using the
$\mct2^{\textrm{\textrm{max}}}(\chi)$ under realistic $|\dt|$ values
of $\chargino$ pairs arising from squark/gluino decays.

\vskip0.2cm \noindent\textbf{SUSY example} We demonstrate the mass
measurement of $\chargino$ and $\sneutrino$ for a SUSY benchmark
point described in Ref. \cite{susypoint}. In this scenario, 3rd
generation sfermions are very heavy and $\chargino$ has no branching
ratio (BR) to $\stau$ with BR$(\chargino\rightarrow
\ell+\sneutrino)=0.63$ followed by $\sneutrino\rightarrow\lnp+\nu$.
The relevant mass spectrum of the point is
$(m_{\gluino},m_{\squark},m_{\chargino},m_{\sneutrino},m_{\lnp})
=(\sim720,\sim635,231.5,157.2,123.3)\GeV$ with
$\sigma^{SUSY}_{tot}\sim17\pb$. We calculate $\mct2(\chi)$ for a
system of a pair of SS chargino decays, where $\alt$ and $\bet$ are
the transverse momenta of leptons from the chargino decays, and the
two missing $\sneutrino$s' momenta provide $\etmiss$ \cite{foot1}.
All of the other jets in the event are taken as the $\dt$ of the
chargino decay system. SUSY events of $70fb^{-1}$ at
$\sqrt{s}=14\,\TeV$ are generated by \verb"PYTHIA" \cite{pythia} for
this point. The detector simulation is implemented with \verb"PGS4"
\cite{pgs}.
\begin{figure}[t]
\begin{center}
\includegraphics[width=0.5\textwidth]{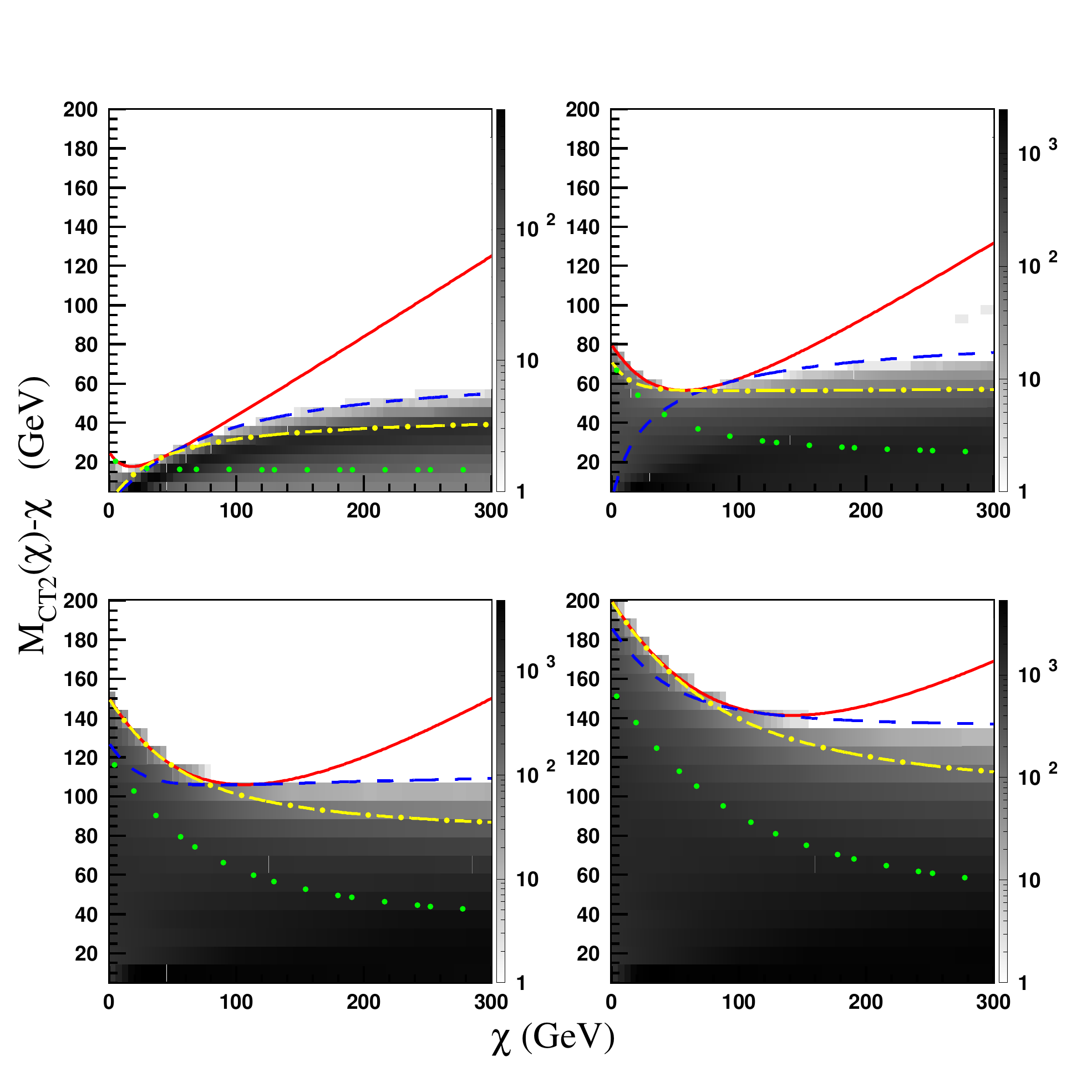}
\caption{Contour plots of the SS SUSY dileptonic events surviving
the event selection cuts in $\mct2(\chi)-\chi~\rm{vs}~\chi$ plane
for
$|\dt^{\textrm{max}}|=(50_{\textrm{Top-Left}},\,160_{\textrm{T-Right}},\,300_{\textrm{Bottom-L}},\,
400_{\textrm{B-R}})\GeV$.}\label{fig:mct2max1}
\end{center}
\end{figure}

We select SS dileptonic events with $N(\geq2)$-jets + $\etmiss$. The
event selection cuts are $P_T(\ell_{1,2})>20\GeV$,
$P_T(jet_{1,2})>(100,80)\GeV$, and $\etmiss>100\GeV$. After the
cuts, the signal to background ratio becomes high and the kinematic
endpoints are expected to appear clearly without background from the
SM events and neutralino decays \cite{atlas2}. Additionally, we
impose a cut on the $|\dt|$ value, $|\dt|\leq|\dt^{\textrm{max}}|$.
For given $|\dt^{\textrm{max}}|$ and $\chi$ values,
$\mct2^{\textrm{max}}(\chi)$ (\ref{mct2left},\ref{mct2right}) of the
signal events is completely described by two known
($|\bar{\delta}_T|=|\dt^{\textrm{max}}|/2,\chi$) and two unknown
($m_{\chargino},m_{\sneutrino}$) parameters. Fig.
(\ref{fig:mct2max1}) shows the contour histogram of the events which
survived the cuts in the $\mct2(\chi)-\chi$ vs $\chi$ plane for
$|\dt^{\textrm{max}}|=(50_{(a)},\,180_{(b)},\,300_{(c)},\,400_{(d)})\GeV$.
The events are populated below the expected endpoints of
$\mct2(\chi)$ (\ref{mct2left},\ref{mct2right}). The solid lines show
$\mct2^{\textrm{max}}(\chi)$ (\ref{mct2left}) for all of the
$|\dt^{\textrm{max}}|$ values. The boundaries of the $\mct2$
distributions are well described by the solid lines in the range of
$\chi\leq\chi_{*}(=53.4_{(a)},91.3_{(b)},109.3_{(c)},120.7_{(d)})\GeV$
in Fig. (\ref{fig:mct2max1}). In each plot, we also show three
functional lines of
$\mct2^{\textrm{max}}(\chi>\chi_*,|\dt^{\textrm{max}}|)$
(\ref{mct2right}) for three different mass spectrums,
$m_{Y(\textrm{Dashed,Dashed-Dotted,Dotted})}=(m_{\chargino},m_{\chargino}-30,m_{\chargino}-60)\GeV$,
with the same $m_{X}=m_{\sneutrino}$ values. The boundary of the
$\mct2(\chi,|\dt^{\textrm{max}}|)$ distributions also appear to be
well-matched with the expected \textrm{Dashed}-lines for
$\chi\geq\chi_*$.
\begin{figure}[t!]
\begin{center}
\includegraphics[width=0.5\textwidth]{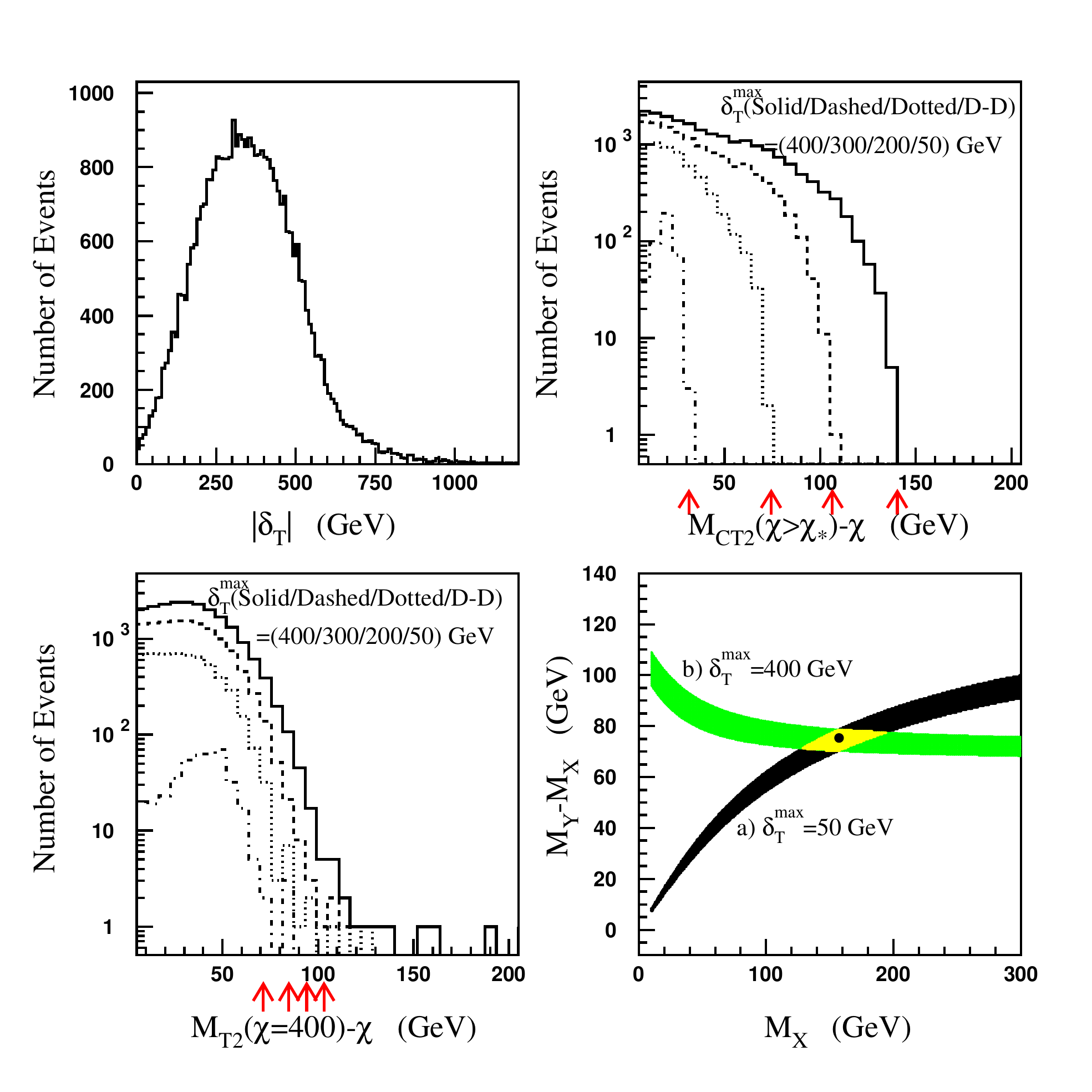}
\caption{ a)$^{\textrm{T-L}}$ $|\dt|$ distribution of the SS
dilepton + $\etmiss$ system surviving the cuts. b)$^{\textrm{T-R}}$
$\mct2$ and c)$^{\textrm{B-L}}$ $\mt2$ distributions.
d)$^{\textrm{B-R}}$ Consistent mass ranges in $M_Y-M_X$ vs $M_X$
plane. }\label{fig:mct2max2}
\end{center}
\end{figure}

Using the reconstructed $\mct2^{\textrm{max}}(\chi)$ for different
$|\dt^{\textrm{max}}|$ values, we finally attempt to get the
consistent range of chargino and sneutrino masses. Fig.
(\ref{fig:mct2max2}(a)) shows the $|\dt|$ distribution
of the SS dilepton + $\etmiss$ system after the cuts. %The first
%peak in low $|\dt|$ region is from direct SS chargino pair
%production and
The event statistics near $|\dt|\sim O(100)\GeV$ are very rich due
to the large boost effect of hard jets from initially produced
colored superparticles, and it can be a good testbed for the mass
measurement utilizing various transverse boost effects. For
different values of
$|\dt^{\textrm{max}}|=(400_{\textrm{Solid}},300_{\textrm{Dashed}},
200_{\textrm{Dotted}},50_{\textrm{D-D}})\GeV$ cuts, corresponding
$\mct2(\chi=\chi_*+\sim20\GeV)$ distributions are shown as in Fig.
(\ref{fig:mct2max2}(b)). Fig. (\ref{fig:mct2max2}(c)) displays
$\mt2(\chi=400\GeV)$ distributions with the $|\dt^{\textrm{max}}|$
cuts. In both of the plots, red arrows denote the expected endpoints
of the $M_{CT2/T2}(\chi)$ for the given $|\dt^{\textrm{max}}|$ cuts.
Then one can clearly see that\vskip 0.09cm
\begin{enumerate}
    \item The endpoint structure of the $\mct2(\chi)$
distribution is statistically cleaner with less pollution
\cite{cmct2} and sharper \cite{mct2} in wide range of
$|\dt^{\textrm{max}}|$, compared with $\mt2(\chi)$.
    \item The $\mct2$ is so sensitive under the transverse boost
effect, the shift of $\mct2^{\textrm{max}}$ with a change of
$|\dt^{\textrm{max}}|$ can be sizable well beyond the uncertainties
in endpoint extraction, as well as the number of events contributing
to
$\mct2(\chi,|\dt|>|\dt^{\textrm{max}}|)>\mct2^{\textrm{max}}(\chi,|\dt^{\textrm{max}}|)$,
is much larger than the number of the events with $M_{T2}(\chi,
|\delta_T|>|\delta_T^{max}|) > M_{T2}^{\textrm{max}}(\chi,
|\delta_T^{max}|)$
\end{enumerate}
\noindent Putting all these
facts together, we now have a collider variable with which the
measurement of Tr. boost-shifted endpoints are experimentally
reliable for mass determination under various systematic
uncertainties of endpoint extraction.

An error profile function of $\chi$,
$\epsilon(\chi)=a\,\setminus\,b(\chi-\chi_*)+c$ is assumed in the
two regions of ($\chi\leq\chi_*\setminus\chi>\chi_*$), which are
likely to be obtained by simple endpoint fitting of
$\mct2(\chi,|\dt^{\textrm{max}}|)$ distributions. The coefficients
are conservatively set to $(a,b,c)=(2|3,0.03,2|3)\GeV$ for
$|\dt^{\textrm{max}}|=(50|400)\GeV$. Then, for the two
$|\dt^{\textrm{max}}|$ values we fit two $\alpha$ values in Eq.
(\ref{2ndkink}) and (\ref{mct2right}), using the parameterized
endpoint measurements, $\mct2^{\textrm{max}}(\chi,|\dt^{\max}|)\pm
\epsilon(\chi)$ in the range of $0\leq \chi \leq 300 \GeV$,
including $\chi_*$ which is also a function of $\alpha$. This is a
one-parameter fitting of $\alpha$ for a given $|\dt^{\textrm{max}}|$
value, and the precision of fit result is not much affected by a
change of the fit region of $\chi$, as long as it is taken broadly
(0,$\chi_*+ O(50-100)\GeV$) enough to include the $\chi_*$ region.
For the parameterized errors, $\alpha\pm \delta \alpha$ values are
fitted to be $(69.5|136.5)\pm(3.38|5.14)\GeV$ for
$|\dt^{\textrm{max}}|=(50|400)\GeV$, with $\chi^2/ndf<1$. Fig.
(\ref{fig:mct2max2}(d)) shows two bands$_{\textrm{(Green|Black)}}$
in $M_Y-M_X~vs~M_X$ plane, consistent with the fitted $\alpha$
values. The black dot indicates the true mass point and the yellow
intersection region is the $1\sigma$ range of the mass measurement
using the $\mct2^{\textrm{max}}(\chi)$ data for two different
$|\dt^{\textrm{max}}|$ cuts. \vskip0.3cm

\noindent\textbf{Discussion} In this paper we study the
$\mct2(\chi)$ endpoint of a system containing a pair of single step
decays, which is expected to be the simplest non-trivial new physics
event topology. The $\mct2$ magnifies the transverse boost effect on
the system, while keeping its endpoint structure clean and sharp.
Thus, measuring $\mct2^{\textrm{max}}$ for different $|\dt|$ values
can provide us with reliable resolving power to determine both
masses of mother and missing daughter particles only using moderate
transverse boost momentum of the system of our interest. With
realistic $\dt$ profiles of \verb"mSUGRA" like events, we
demonstrate simultaneous mass measurement of $\chargino$ and
$\sneutrino$ in SS dileptonic decay channel, where
$\chargino\rightarrow\sneutrino+\ell$. The technique also works for
a single mother particle decay system with non-zero transverse
momentum.

This work was supported by World Premier International Center
Initiative (WPI Program), MEXT, Japan.


\begin{thebibliography}{99}
\bibitem{foot1}
The other decay channel, $\chargino\rightarrow\slepton+\nu_{\ell}$,
is also open, however it does not give the maximum of $\mct2(\chi)$
in this spectrum \cite{cmct2}.
\end{thebibliography}
\end{document}